# Size effects in the electronic properties of finite arrays of exchange coupled quantum dots: A renormalization group approach

J. X. Wang and Sabre Kais

*Department of Chemistry, Purdue University West Lafayette, IN 47907*

F. Remacle [*] and R. D. Levine [†]

*The Fritz Haber Research Center for Molecular Dynamics, The Hebrew University, 91904 Jerusalem, Israel*

Transport properties of arrays of metallic quantum dots are governed by the distance-dependent exchange coupling between the dots. It is shown that the effective value of the exchange coupling, as measured by the charging energy per dot, depends monotonically on the size of the array. The effect saturates for hexagonal arrays of over $7^5$ unit cells. The discussion uses a multi-stage block renormalization group approach applied to the Hubbard Hamiltonian. A first order phase transition occurs upon compression of the lattice and the size dependence is qualitatively different for the two phases.

Two-dimensional arrays with tailored electronic properties are generating much current experimental and theoretical interest [1–6]. This was made possible by the development of synthetic methods for the preparation of, so-called, quantum dots (QDs). There is much current interest in the ability to engineer nanoscale electronic devices, including quantum computers from quantum dots. Quantum dots of nearly identical sizes self assemble into a planar array. (The dots are passivate against collapse by coating them with organic ligands). For, e.g., Ag nanodots, the packing is hexagonal. The lower lying electronic states of an isolated dot are discrete being determined by the confining potential (and therefore the size) of the dot. Because of their larger size (100-1000 atoms each) it takes only a relatively low energy to add another electron to a dot, as revealed by scanning tunneling microscopy [7]. This energy is much lower than the corresponding energy for ordinary atoms and most molecules. It follows that when dots are close enough to be exchange coupled, which is the case in an array, the charging energy can be quite low. We here propose and implement a computational method that allows the contributions of such ionic configurations even for extended arrays. The technical problem is that the Coulombic repulsion between two electrons (of opposite spins) that occupy the same dot cannot be described in a one electron approximation. It requires allowing for correlation of electrons. Most methods that explicitly include correlation effects scale as some high power of the number of atoms (here, dots) and are computationally intractable. For example, a hexagonal array of only 19 dots, 3 dots per side, has already 2,891,056,160 low electronic configurations. So earlier [8] exact computations including charging energy were limited to a hexagonal array of only 7 dots, 2 dots per side. Yet current measurements of both static [9] and transport [10] properties use arrays of at least 100 dots per side. The simplest Hamiltonian that includes both the Coulombic (or charging energy) effects and the exchange coupling is the Hubbard model [11]. This model can be solved exactly for a one-dimensional chain, but for a two-dimensional array it is, so far, analytically intractable. In the absence of a closed solution, various methods have been developed [12]. Renormalization group(RG) methods are receiving increasing attention because of their non-perturbative nature, that allows application to the intermediate-to-strong coupling regime. This Letter reports the applications of a real-space block renormalization group (BRG) method [13–15] on a 2-dimensional triangular lattice with hexagonal blocks as shown in Fig. 1. This enables us to explore the evolution of size effects. Specifically, we demonstrate how the effective exchange coupling between adjacent dots is modulated by the size of the array. In other words, we argue that for finite arrays of quantum dots, their response will be size dependent and we determine the scaling.

[*]Maître de Recherches, FNRS, Belgium. Permanent address : Département de Chimie, B6c, Université de Liège, B4000 Liège, Belgium.
[†]Corresponding author : email rafi@fh.huji.ac.il, fax : 972-2-6513742

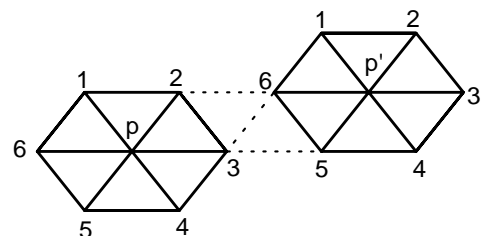

FIG. 1: Schematic diagram of the triangular lattice with hexagonal blocks. Only two neighboring blocks $p$ and $p'$ are drawn here. The dotted lines represent the interblock interactions and solid line intrablock ones.

The essential physics of the collective low energy elec-tronic structure of an array of metallic quantum dots has the following ingredients. Each dot is characterized by the energy $\mu$ needed to remove its highermost (=valence) electron. Adjacent dots are exchange coupled with a strength, $t$, that decreases exponentially at larger distances but saturates at very close packing. Transport processes are facilitated by the coupling and are therefore measured in the intermediate to strong coupling range, $U/t$. When $t$ is large it can overwhelm the role of the charging energy $U$ and it is then possible to treat the lattice by a tight binding approximation that can be easily applied to larger size arrays. However, over much of the experimentally available compression range it is the case that $t < U$. Our letter specifically addresses this intermediate coupling range where the correlation of electrons needs to be explicitly accounted for.

The Hubbard Hamiltonian is written as[16]

$$H = -t \sum_{<i,j>,\sigma} c^{\dagger}_{i\sigma} c_{j\sigma} + U \sum_i n_{i\uparrow} n_{i\downarrow} - \mu \sum_i (n_{i\uparrow} + n_{i\downarrow}) \quad (1)$$

$c^{\dagger}_{i\sigma}(c_{i\sigma})$ creates(annihilates) an electron with spin $\sigma$ in the valence (=Wannier) orbital of the dot located at site $i$; the corresponding number operator is $n_{i\sigma} = c^{\dagger}_{i\sigma} c_{i\sigma}$. The angular bracket $< ... >$ on the first sum in Eq. (1) indicates that summation is restricted to nearest-neighbor dots. This model Hamiltonian allows only one orbital per dot. That orbital can be empty or it can accommodate one or two electrons.

The essence of the BRG method is to map the above many-particle Hamiltonian on a lattice to a new one with fewer degrees of freedom and with the same low-lying energy levels [17]. Then the mapping is repeated leading to a final Hamiltonian of a seven site hexagonal array for which we obtain an exact numerical solution. The procedure can be summarized into three steps: First divide the $N$–site lattice into appropriate $n_s$–site blocks labeled by $p(p=1,2,..., N/n_s)$ and separate the Hamiltonian $H$ into intrablock part $H_B$ and interblock $H_{IB}$:

$$H = H_B + H_{IB} = \sum_p H_p + \sum_{\langle p,p'\rangle} V_{p,p'} \quad (2)$$

where $H_p$ is the Hamiltonian (1) for a given block and the interblock $p,p'$ coupling is defined in (3) below.

The second step is to solve $H_p$ exactly for the eigenvalues $E_{p_i}$ and eigenfunctions $\Phi_{pi}(i=1,2,...,4^{n_s})$. Then the eigenfunctions of $H_B$ are constructed by direct multiplication of $\Phi_{pi}$. The last step is to treat each block as one site on a new lattice and the correlations between blocks as hopping interactions.

The original Hilbert space has four states per site. If we are only concerned with lower lying states of the system as when studying the metal-insulator-transition [18], it is not necessary to keep all the states for a block.

To make the new Hamiltonian tractable the reduction in size should not be accompanied by a proliferation of new couplings. Then one can use an iteration procedure to solve the model. To achieve this it is necessary to keep only 4 states in step 2. Their energies are $E_i(i=1,2,3,4)$. In order to avoid proliferation of additional couplings in the new Hamiltonian the four states kept from the block cannot be arbitrarily chosen. Some definite conditions as discussed in Ref. [16] must be satisfied. For example, the states must belong to the same irreducible representation of $C_{6\nu}$ symmetry group of the lattice. In particular, in order to copy the intrasite structure of the old Hamiltonian, one must have that $E_3 = E_4$. Furthermore, particle-hole symmetry of a half-filled lattice requires that $E_1 = E_2$. Further restrictions follow from the need to make extra couplings vanish. Operators in the truncated basis are denoted by a prime so that the interblock coupling of Eq. (2) is

$$V_{pp'} = \nu\lambda^2 t \sum_{\sigma} c'^{+}_{p\sigma} c'_{p'\sigma}, \quad (3)$$

where $\nu$ represents the number of couplings between neighboring blocks. The coupling strength for the border sites of a block by $\lambda$ and the renormalization group equation for the coupling strength is

$$t' = \nu\lambda^2 t, \quad (4)$$

The other renormalization relations are

$$U' = 2(E_1 - E_2), \quad K' = (E_1 + E_2)/2. \quad (5)$$

$K$ sets the zero of energy with the initial value of $K = -U/4$. For the half-filled lattice $\mu = U/2$. Below we will compute the energy gap for adding or removing an electron from a half-full lattice. The gap is independent of the choice for $\mu$. The results of the procedure are applied to a finite array of $7^n$ sites in Fig. 2. The coupling constants for the initial array are $U = U_0$ and $t_0 = 1$ so that the abscissa spans the intermediate coupling range $U_0 > t_0$. The ordinate gives the renormalized coupling constants, $U'/t'$, on a logarithmic scale, for various values of the size, $n$, of the original lattice ($n = 1$ being shown for reference because it is the line $U'/t' = U_0/t_0$). There are two features to note. One is that there is a phase transition at $U_0/t_0 = 12.5$. This is quite near the recent result $U_0/t_0 = 12.07$ obtained with exact diagonalization method [19]. The other feature of the results is the one we want to focus on. It is the variation of the renormalized coupling constants with size of the array. The reason for drawing attention to this scaling is as follows.

The low lying electronic states of the large array are computed by an exact diagonalization [8] of the Hubbard-like Hamiltonian of a seven site array, the smallest hexagonal structure, with the renormalized coupling constants. In the (albeit, approximate) real space group renormalization procedure that we use the only way that the size of the original array comes in is in the value of the renormalized coupling constants. Arrays of all sizes can therefore be made to fall on a common universal plot where they differ only in the input values of the coupling constants. To be sure, the actual value of the coupling constant $U'/t'$ depends on $n$, and quite systematically and dramatically so, Fig. 3, but for given initial coupling constant, $U_0/t_0$, there is no other dependence. In our earlier work we have argued, on the basis of perturbation theory and also exact computations that in the regime $t > U$ the effects of the charging energy can be neglected, particularly so in application to larger arrays [6, 9, 20]. Fig. 3 indicates that this should be the case anywhere to the left of the phase transition, where $U'/t'$ decreases strongly with increasing $n$. As discussed below, the low lying electronic states are determined by the value of $U'/t'$ and therefore, in the intermediate to strong coupling regime the importance of the dot-dot exchange coupling increases with the size of the lattice. Conversely so in the weak coupling regime. To emphasize the universal scaling we discuss the gap $\Delta$ between the highest occupied and lowest empty state of the entire array. The closing of this gap is the signature of the Mott insulator to metal transition. We define the gap as usual by the energy cost difference between adding or removing one electron from the half-filled array. This gap is the limiting value of $U'$ after a large number of iterations but it can also be computed for any finite value of lattice size $n$. As is to be expected, such a plot of the scaled gap, $\Delta_n/t'$ vs. $U_0$ looks remarkably similar to Fig. 2 that showed the scaled coupling constant, $U'/t'$ vs. $U_0$. In particular, such a plot establishes the phase transition at $U_0/t_0 = 12.5$. Here we present it as a universal plot, Fig. 4, showing the scaled gap vs. the scaled coupling constant. The results for the gap for different values of $n$ (identified in the legend) all fall on the same curve.

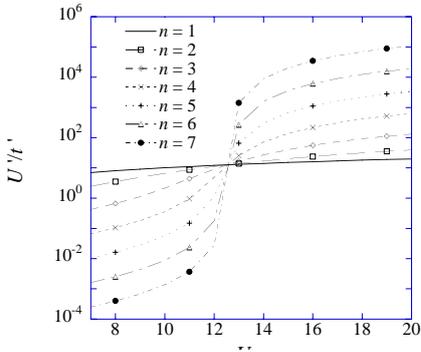

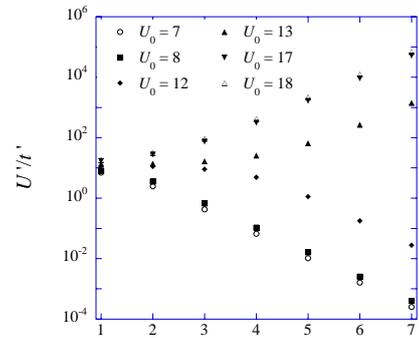

FIG. 2: The renormalized coupling strength, $U'/t'$, logarithmic scale, starting from a lattice of $n$ hexagons of seven dots each, vs. the initial value of the charging energy $U_0/t_0$, (measured in units of the exchange coupling $t_0$ so that the renormalization equations begin with $t_0 = 1$). The solid line is that for a single hexagon for which $U'/t' = U_0/t_0$. The results show that there is a phase transition at $U_0/t_0 = 12.5$. Note how the size effect is different at the two sides of the transition and how the role of the exchange coupling is increasingly important for compressed ($U_0/t_0$ smaller) lattices.

FIG. 3: The renormalized coupling strength, $U'/t'$, logarithmic scale, for a lattice of $n$ hexagons of seven dots each, vs. $n$ for different values of the charging energy $U_0$ of the full lattice. Note the qualitatively different trends in the weak and the intermediate coupling regimes.

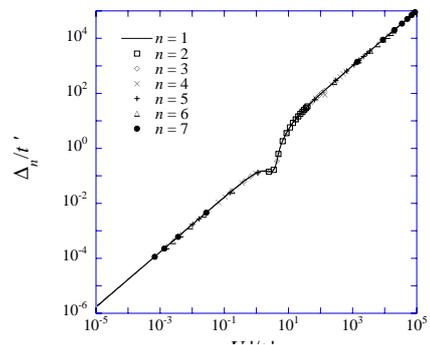

FIG. 4: The gap (=difference between the energy needed to add or to remove an electron from a half-filled lattice) for a lattice of $n$ hexagons, logarithmic scale, vs. the renormalized coupling strength. The discrete points are computed for different values of $n$ and charging energy $U_0$ of the full lattice. The line is the exactly computed value of the gap for a single hexagon, for which $U'/t' = U_0/t_0$. The kink is the remnant of the phase transition for $n \to \infty$.

The line going through the individual points is not a fit but is the results of exact computations of the gap for a hexagonal array of seven dots, $n = 1$. Larger arrays are renormalized first and then the results are put in as discrete points, corresponding to the series of values of $U_0$ shown in Figs. 2 and 3. The kink in the plot is the memory of the phase transition for $n \to \infty$. The points for $n > 1$ are all for the same common set of values of $U_0$ and their spread on this plot is another manifestation of the size effect. Specifically note how points for larger values of $n$ congregate toward the two extremes of the plot.

Fig. 4. shows that the value of $U_0/t_0$ for which the transition occurs depends on the size of the lattice. The value ($U_0/t_0 = 1.5$) is quite low for $n = 1$ and increases with $n$ saturating at 12.5 for $n \to \infty$. The physics of this increase is that originally made by Mott [21] in identifying the insulator to metal transition. It is a competition

between localization, measured by $U$, and delocalization, measured by $t$. Other things being equal, the bigger the lattice the more an electron can delocalize. From the point of view of renormalization, at each iteration we lump seven dots into one. Such a bigger dot has a lower charging energy and furthermore a renormalized dot is more strongly coupled to its neighbors, cf. Eq.(4). Both effects reduce the critical value of $U_0/t_0$. It should be mentioned here that in order to increase the accuracy, the hexagonal blocks instead of the most-used triangular ones are utilized. In fact, for triangular blocks, similar scalings are also found. There is no qualitative difference, which gives us much confidence upon the consistence of the results between the RG procedures with difference kind of blocks. Nanoscale devices are finite arrays of quantum dots. The results of the renormalization group approach suggest that such arrays can be usefully treated and discussed by renormalizing the effective coupling strength down to the unit cell of the array. The scaling with size is predicted to vary depending on the coupling regime. The dot-dot exchange coupling becomes of lesser importance for larger arrays that are weakly coupled (or have a higher charging energy) and it is of increasing effective strength for intermediate to strong coupling. The available computational and experimental evidence [6, 9, 22] is that the coupling regimes and the resulting phase diagram are far richer than the single phase transition that we here discussed. This is for two reasons. One is that there are additional control variables (e.g., external electrical field [9], temperature [6]) that we have not introduced here. The other is the inherent size fluctuations of quantum dots [22] that can induce a transition to a domain localized phase [23]. It corresponds to charge transfer between not near neighboring dots and is similar to the super-exchange coupling in molecular charge transfer. This transition where the electronic states are not localized but are neither extensively spread out seems to have been experimentally detected [9]. We plan to extend the RGB procedure so as to allow us to treat the role of disorder. Preliminary results already verify that this second transition can be seen by renormalization of larger lattices. Equally we seek additional experimental signatures of the variation of the effective coupling strength with the size of the array. In conclusion we note that a single quantum dot is made up of tens to hundreds of atoms (or molecules). Yet its very low lying electronic states can be simply and usefully described by the idea of a quantum confinement. Here we tried to do the same for a finite array of quantum dots. A larger number of hexagons was scaled down to a single hexagon with an effective coupling. This scaling is achieved by collapsing seven dots into one at each stage of the iteration.

One of us (S.K.) would like to acknowledge the financial support of ONR and NSF.